\documentclass[aps,prd,longbibliography,reprint,twocolumn,amsmath,amssymb,amsfonts,showpacs,footnote]{revtex4-1}%reprint ,superscriptaddress

\usepackage{ifpdf}
\usepackage{dcolumn}
\usepackage{enumitem}
\usepackage{bm}
\usepackage{here}
\ifpdf

      \usepackage[pdftex]{graphicx}  % note the x at the end
      \usepackage[pdftex]{epsfig}

      \usepackage[pdftex]{hyperref}
\hypersetup
	{
		colorlinks,%
		citecolor=blue,%
		linkcolor=red,%
		urlcolor=blue,%
	}

\else

      \usepackage[dvips]{graphicx}  % note the x at the end
      \usepackage[dvips]{epsfig}

      \usepackage[dvipdfm]{hyperref}
      \hypersetup
	{
		colorlinks,%
		citecolor=blue,%
		linkcolor=red,%
		urlcolor=blue,%
	}
\fi

\begin{document}

\title{Waveforms in massive gravity and neutralization of giant black hole ringings}

\author{Yves D\'ecanini}
\email{decanini@univ-corse.fr}
\affiliation{Equipe Physique
Th\'eorique - Projet COMPA, \\ SPE, UMR 6134 du CNRS
et de l'Universit\'e de Corse,\\
Universit\'e de Corse, Facult\'e des Sciences, BP 52, F-20250 Corte,
France}

\author{Antoine Folacci}
\email{folacci@univ-corse.fr}
\affiliation{Equipe Physique
Th\'eorique - Projet COMPA, \\ SPE, UMR 6134 du CNRS
et de l'Universit\'e de Corse,\\
Universit\'e de Corse, Facult\'e des Sciences, BP 52, F-20250 Corte,
France}

\author{Mohamed {Ould El Hadj}}
\email{ould-el-hadj@univ-corse.fr}
\affiliation{Equipe Physique
Th\'eorique - Projet COMPA, \\ SPE, UMR 6134 du CNRS
et de l'Universit\'e de Corse,\\
Universit\'e de Corse, Facult\'e des Sciences, BP 52, F-20250 Corte,
France}

\date{\today}

\begin{abstract}

A distorted black hole radiates gravitational waves in order to settle down in a smoother geometry. During that relaxation phase, a characteristic damped ringing is generated. It can be theoretically constructed from both the black hole quasinormal frequencies (which govern its oscillating behavior and its decay) and the associated excitation factors (which determine intrinsically its amplitude) by carefully taking into account the source of the distortion. In the framework of massive gravity, the excitation factors of the Schwarzschild black hole have an unexpected strong resonant behavior which, theoretically, could lead to giant and slowly decaying ringings. If massive gravity is relevant to physics, one can hope to observe these extraordinary ringings by using the next generations of gravitational wave detectors. Indeed, they could be generated by supermassive black holes if the graviton mass is not too small. In fact, by focusing on the odd-parity $\ell=1$ mode of the Fierz-Pauli field, we shall show here that such ringings are neutralized in waveforms due to (i) the excitation of the quasibound states of the black hole and (ii) the evanescent nature of the particular partial modes which could excite the concerned quasinormal modes. Despite this, with observational consequences in mind, it is interesting to note that the waveform amplitude is nevertheless rather pronounced and slowly decaying (this effect is now due to the long-lived quasibound states). It is worth noting also that, for very low values of the graviton mass (corresponding to the weak instability regime for the black hole), the waveform is now very clean and dominated by an ordinary ringing which could be used as a signature of massive gravity.

\end{abstract}

\pacs{04.70.Bw, 04.30.-w, 04.25.Nx, 04.50.Kd}

\maketitle

\section{Introduction}

In a recent article \cite{Decanini:2014bwa} (see also the preliminary note \cite{Decanini:2014kha}), we have discussed a new and unexpected effect in black hole (BH) physics: for massive bosonic fields in the Schwarzschild spacetime, the excitation factors of the quasinormal modes (QNMs) have a strong resonant behavior around critical values of the mass parameter leading to giant ringings which are, in addition, slowly decaying due to the long-lived character of the QNMs. We have described and analyzed this effect numerically and confirmed it analytically by semiclassical considerations based on the properties of the unstable circular geodesics on which a massive particle can orbit the BH. We have also focused on this effect for the massive spin-$2$ field. Here, we refer to Refs.~\cite{Hinterbichler:2011tt,deRham:2014zqa} for recent reviews on massive gravity, to Refs.~\cite{Volkov:2013roa,Babichev:2015xha} for reviews on BH solutions in massive gravity, and to Refs.~\cite{Babichev:2013una,Brito:2013wya,Brito:2013yxa,Hod:2013dka,deRham:2014zqa,Babichev:2015xha,Babichev:2015zub} for articles dealing with gravitational radiation from BHs and BH perturbations in the context of massive gravity.

In our previous works \cite{Decanini:2014bwa,Decanini:2014kha}, we have considered the Fierz-Pauli theory in the Schwarzschild spacetime \cite{Brito:2013wya} which can be obtained by linearization of the ghost-free bimetric theory of Hassan, Schmidt-May, and von Strauss discussed in Ref.~\cite{Hassan:2012wr} and which is inspired by the fundamental work of de Rham, Gabadadze, and Tolley \cite{deRham:2010ik,deRham:2010kj}. For this spin-$2$ field, we have considered more particularly the odd-parity $(\ell=1,n=0)$ QNM. (Note that it is natural to think that similar results can be obtained for all the other QNMs -- see also Ref.~\cite{Decanini:2014bwa}.) We have then shown that the resonant behavior of the associated excitation factor occurs in a large domain around a critical value ${\tilde \alpha}_0\approx 0.90$ of the dimensionless mass parameter ${\tilde \alpha}=2M\mu /{m_\mathrm{P}}^2$ (here $M$, $\mu$ and $m_\mathrm{P}= \sqrt{\hbar c /G} $ denote, respectively, the mass of the BH, the rest mass of the graviton, and the Planck mass) where the QNM is weakly damped. It is necessary to recall that the Schwarzschild BH interacting with a massive spin-$2$ field is, in general, unstable \cite{Babichev:2013una,Brito:2013wya} (see, however, Ref.~\cite{Brito:2013yxa}). In the context of the massive spin-$2$ field theory we consider, this instability is due to the behavior of the (spherically symmetric) propagating $\ell=0$ mode \cite{Brito:2013wya}. It is, however, important to note that:

\begin{enumerate}[label=(\roman*)]
\item   It is a ``low-mass'' instability which disappears above a threshold value ${\tilde \alpha}_t \approx 0.86$ of the reduced mass parameter ${\tilde \alpha}$ and that the critical value around which the quasinormal resonant behavior occurs lies in the stability domain, i.e., ${\tilde \alpha}_0 > {\tilde \alpha}_t$.
    
\item  Even if a part of the $\tilde\alpha$ domain where the quasinormal resonant behavior occurs lies outside the stability domain (i.e., below ${\tilde\alpha}_t$), one can nevertheless consider the corresponding values of the reduced mass parameter; indeed, for graviton mass of the order of the Hubble scale, the instability timescale is of order of the Hubble time and the BH instability is harmless.
\end{enumerate}

\noindent As a consequence, the slowly decaying giant ringings predicted in the context of massive gravity seem physically relevant (they could be generated by supermassive BHs -- see also the final remark in the conclusion of Ref.~\cite{Decanini:2014kha}) and could lead to fascinating observational consequences which could be highlighted by the next generations of gravitational wave detectors.

In the present article, by assuming that the BH perturbation is generated by an initial value problem with Gaussian initial data (we shall discuss, in the conclusion, the limitation of this first hypothesis), an approach which has regularly provided interesting results (see, e.g., Refs.~\cite{Leaver:1986gd,Andersson:1996cm,Berti:2006wq}), and by restricting our study to the odd-parity $\ell=1$ mode of the Fierz-Pauli theory in the Schwarzschild spacetime (we shall come back, in the conclusion, on this second hypothesis) but by considering the full signal generated by the perturbation and not just the purely quasinormal contribution, we shall show that, in fact, the extraordinary BH ringings are neutralized in waveforms due to the coexistence of two phenomena:

\begin{enumerate}[label=(\roman*)]

  \item The excitation of the quasibound states (QBSs) of the Schwarzschild BH. Indeed, it is well known that, for massive fields, the resonance spectrum of a BH includes, in addition to the complex frequencies associated with QNMs, those corresponding to QBSs. Here, we refer to Refs.~\cite{Deruelle:1974zy,Damour:1976kh,Zouros:1979iw,Detweiler:1980uk} for important pioneering works on this topic and to Refs.~\cite{Brito:2013wya,Babichev:2015xha} for recent articles dealing with the QBS of BHs in massive gravity. In a previous article \cite{Decanini:2015yba}, we have considered the role of QBSs in connection with gravitational radiation from BHs. By using a toy model in which the graviton field is replaced with a massive scalar field linearly coupled to a plunging particle, we have highlighted in particular that, in waveforms, the excitation of QBSs blurs the QNM contribution. Unfortunately, due to numerical instabilities, we have limited our study to the low-mass regime. Now, we are able to overcome these numerical difficulties and we shall observe that, near the critical mass ${\tilde \alpha}_0$, the QBSs of the BH not only blur the QNM contribution but provide the main contribution to waveforms.

  \item The evanescent nature of the particular partial mode which could excite the concerned QNM and generate the resonant behavior of its associated excitation factor. Indeed, if the mass parameter lies near the critical value ${\tilde \alpha}_0$, we shall show that the real part of the quasinormal frequency is smaller than the mass parameter and lies into the cut of the retarded Green function. In other words, the QNM is excited by an evanescent partial mode and, as a consequence, this leads to a significant attenuation of its amplitude.

\end{enumerate}

\noindent It is interesting to note that, despite the neutralization process, the waveform amplitude remains rather pronounced (if we compare it with those generated in the framework of Einstein's general relativity) and slowly decaying, this last effect being now due to the excited long-lived QBSs.

In the article, even if it was not our main initial concern, we have also briefly consider the behavior of the waveform for very small values of the reduced mass parameter $\tilde \alpha$ corresponding to the weak instability regime. Indeed, our results concerning the QNMs as well as the QBSs of the Schwarzschild BH have permitted us to realize that the waveform associated with the odd-parity $\ell=1$ mode of the Fierz-Pauli theory could be helpful to test massive gravity even if the graviton mass is very small: the fundamental QNM generates a ringing which is neither giant nor slowly decaying but which is not blurred by the QBS contribution.

Throughout this article, we adopt units such that $\hbar = c = G = 1$. We consider the exterior of the Schwarzschild BH of
mass $M$ defined by the metric $ds^2= -(1-2M/r)dt^2+ (1-2M/r)^{-1}dr^2+ r^2 d\sigma_2^2$ (here $d\sigma_2^2$ denotes the metric
on the unit $2$-sphere $S^2$) with the Schwarzschild coordinates $(t,r)$ which satisfy $t \in ]-\infty,
+\infty[$ and $r \in ]2M,+\infty[$. We also use the so-called tortoise coordinate $r_\ast \in ]-\infty,+\infty[$ defined from the
radial Schwarzschild coordinate $r$ by $dr/dr_\ast=(1-2M/r)$ and given by $r_\ast(r)=r+2M \ln[r/(2M)-1]$ and assume a harmonic time dependence $\exp(-i\omega t)$ for the spin-$2$ field.

\section{Waveforms generated by an initial value problem and neutralization of giant ringings}
\label{Sec.II}

\subsection{Theoretical considerations}

\subsubsection{Construction of the waveform}

We consider the massive spin-$2$ field in the Schwarzschild spacetime and we focus on the odd-parity $\ell=1$ mode of this field theory (see Ref.~\cite{Brito:2013wya}). The corresponding partial amplitude $\phi (t,r)$ satisfies (to simplify the notation, the angular momentum index $\ell=1 $ will be, from now on, suppressed in all formulas)
\begin{equation}\label{Phi_ell1}
\left[-\frac{\partial^2 }{\partial t^2}+\frac{\partial^2}{\partial r_\ast^2}-V(r)  \right] \phi (t,r)=0
\end{equation}
with the effective potential $V(r)$ given by
\begin{equation}\label{pot_RW_Schw}
V(r) = \left(1-\frac{2M}{r} \right) \left(\mu^2+
\frac{6}{r^2} -\frac{16M}{r^3}\right).
\end{equation}
We describe the source of the BH perturbation by an initial value problem with Gaussian initial data. More precisely, we consider that the partial amplitude $\phi (t,r)$ is given, at $t=0$, by $\phi (t=0,r)=\phi_0(r)$ with
\begin{equation}\label{Cauchy_data}
\phi_0(r)= \phi_0 \exp \left[-\frac{a^2}{(2M)^2} (r_\ast(r) - r_\ast(r_0) )^2 \right]
\end{equation}
and satisfies $\partial_t\phi (t=0,r)=0$. By Green's theorem, we can show that the time evolution of $\phi (t,r)$ is described, for $t>0$, by
\begin{equation}\label{TimeEvolution}
\phi (t,r)=\int_{-\infty}^{+\infty} \partial_t G_\mathrm{ret}(t;r,r') \phi_0(r')   dr'_\ast.
\end{equation}
Here we have introduced the retarded Green function $G_\mathrm{ret}(t;r,r')$ solution of
\begin{equation}\label{Gret}
\left[-\frac{\partial^2 }{\partial t^2}+\frac{\partial^2}{\partial r_\ast^2}-V(r)  \right] G_\mathrm{ret}(t;r,r')=-\delta (t)\delta (r_\ast-r_\ast')
\end{equation}
\noindent and satisfying the condition $G_\mathrm{ret}(t;r,r') = 0$ for $t \le 0$. We recall that it can be written as
\begin{equation}\label{Gret_om}
G_\mathrm{ret}(t;r,r')=-\int_{-\infty +ic}^{+\infty +ic}  \frac{d\omega}{2\pi}  \frac{\phi^\mathrm{in}_{\omega}(r_<) \phi^\mathrm{up}_{\omega}(r_>)}{W (\omega)} e^{-i\omega t}
\end{equation}
\noindent where $c>0$, $r_< =\mathrm{min} (r,r')$, $r_> =\mathrm{max} (r,r')$ and with $W (\omega)$ denoting the Wronskian of the functions $\phi^\mathrm{in}_{\omega}$ and $\phi^\mathrm{up}_{\omega}$. These two functions are linearly independent solutions of the Regge-Wheeler equation
\begin{equation}\label{RW}
\frac{d^2 \phi_{\omega}}{dr_\ast^2} + \left[ \omega^2 -V(r)\right]  \phi_{\omega}=0.
\end{equation}
When $\mathrm{Im} (\omega) > 0$, $\phi^\mathrm{in}_{\omega}$ is uniquely defined by its ingoing behavior at the event horizon $r=2M$ (i.e., for $r_\ast \to -\infty$)
\begin{subequations}\label{bc_in}
\begin{equation}\label{bc_1_in}
\phi^\mathrm{in}_{\omega} (r) \underset{r_\ast \to -\infty}{\sim} e^{-i\omega r_\ast}
\end{equation}
and, at spatial infinity $r \to +\infty$ (i.e., for $r_\ast \to +\infty$), it has an
asymptotic behavior of the form
\begin{eqnarray}\label{bc_2_in}
& & \phi^\mathrm{in}_{\omega}(r) \underset{r_\ast \to +\infty}{\sim}
 \left[ \frac{\omega}{p(\omega)}
\right]^{1/2}  \nonumber \\
& & \quad \times \left(A^{(-)} (\omega) e^{-i[p(\omega)
r_\ast + [M\mu^2/p(\omega)] \ln(r/M)]}\right. \nonumber \\
& & \quad \quad  \left. + A^{(+)} (\omega) e^{+i[p(\omega) r_\ast +
[M\mu^2/p(\omega)] \ln(r/M)]} \right).
\end{eqnarray}
\end{subequations}
Similarly, $\phi^\mathrm{up}_{\omega}$ is uniquely defined by its outgoing behavior at spatial infinity
\begin{subequations}\label{bc_up}
\begin{equation}\label{bc_1_up}
\phi^\mathrm{up}_{\omega} (r) \underset{r_\ast \to +\infty}{\sim}  \left[ \frac{\omega}{p(\omega)}
\right]^{1/2} e^{+i[p(\omega) r_\ast +
[M\mu^2/p(\omega)] \ln(r/M)]}
\end{equation}
and, at the horizon, it has an asymptotic behavior of the form
\begin{equation}\label{bc_2_up}
\phi^\mathrm{up}_{\omega}(r) \underset{r_\ast \to -\infty}{\sim}
B^{(-)} (\omega) e^{-i\omega r_\ast}  + B^{(+)} (\omega) e^{+i\omega r_\ast}.
\end{equation}
\end{subequations}
In Eqs.~(\ref{bc_in}) and (\ref{bc_up}),

\begin{equation}
\label{p_omega}
p(\omega)=\left( \omega^2 - \mu^2 \right)^{1/2}
\end{equation}

\noindent denotes the ``wave number,'' while $A^{(-)} (\omega)$, $A^{(+)} (\omega)$, $B^{(-)} (\omega)$, and $B^{(+)} (\omega)$ are complex amplitudes which, like the $\mathrm{in}$- and $\mathrm{up}$- modes, can be defined by analytic continuation in the full complex $\omega$ plane or, more precisely, in an appropriate Riemann surface taking into account the cuts associated with the functions $p(\omega)$ and $[\omega/p(\omega)]^{1/2}$. By evaluating the Wronskian $W (\omega)$ at $r_\ast \to -\infty$ and $r_\ast \to +\infty$, we obtain
\begin{equation}\label{Well}
W (\omega) =2i\omega A^{(-)} (\omega) = 2i\omega B^{(+)} (\omega).
\end{equation}

Using  (\ref{Gret_om}) into (\ref{TimeEvolution}) and assuming that the source $\phi_0(r)$ given by (\ref{Cauchy_data}) is strongly localized near $r = r_0$ (this can be easily achieved if we assume that the width of the Gaussian function is not too large, i.e., if $a$ is not too small) while the observer is located at a rather large distance from the source, we obtain

\begin{multline}
\label{reponse_partielle}
\phi(t,r)=-\frac{1}{2\pi} \text{Re}\left[\int_{0+i c}^{+\infty+i c}d\omega\left(\frac{e^{-i \omega t}}{A^{(-)}(\omega)}\right)\right. \\
\left.\times \phi_{\omega}^{\text{up}}(r)\int_{-\infty}^{+\infty}dr'_{*}\phi_{0}(r')\phi_{\omega}^{\text{in}}(r')\right].
\end{multline}

\noindent This formula will permit us to construct numerically the waveform for an observer at $(t,r)$.

\subsubsection{Extraction of the QNM contribution}

The zeros of the Wronskian $W (\omega)$ are the resonances of the BH. Here, it is worth recalling that if $W (\omega)$ vanishes, the functions $\phi^\mathrm{in}_{\omega}$ and $\phi^\mathrm{up}_{\omega}$ are linearly dependent. The zeros of the Wronskian lying in the lower part of the first Riemann sheet associated with the function $p(\omega)$ (see Fig.~16 in Ref.~\cite{Decanini:2015yba}) are the complex frequencies of the $\ell=1$ QNMs. Their spectrum is symmetric with respect to the imaginary $\omega$ axis. Similarly, the zeros of the Wronskian lying in the lower part of the second Riemann sheet associated with the function $p(\omega)$ are the complex frequencies of the $\ell=1$ QBSs and their spectrum is symmetric with respect to the imaginary $\omega$ axis.

The contour of integration in Eq.~(\ref{reponse_partielle}) may be deformed in order to capture the QNM contribution  \cite{Leaver:1986gd}, i.e., the extrinsic ringing of the BH. By Cauchy's theorem and if we do not take into account all the other contributions (those arising from the arcs at $|\omega|=\infty$, from the various cuts and from the complex frequencies of the QBSs), we can extract a residue series over the quasinormal frequencies $\omega_n$ lying in the fourth quadrant of the first Riemann sheet associated with the function $p(\omega)$. We then isolate the BH ringing generated by the initial data. It is given by

\begin{equation}
\label{TimeEvolution_QNM}
\phi^\mathrm{QNM} (t,r)= 2 \, \mathrm{Re} \left[ \sum_n i\omega_n {\cal C}_n  e^{-i\omega_n t} \left[\frac{p(\omega_n)}{\omega_n}\right]^{1/2} \!\!\!\!\!\phi_{\omega_n}^\text{up}(r)\right].
\end{equation}

\noindent In this sum, $n=0$ corresponds to the fundamental QNM (i.e., the least damped one) and $n=1,2,\dots $ to the overtones. Moreover, ${\cal C}_n$ denotes the excitation coefficient of the QNM with overtone index $n$. It is defined from the corresponding excitation factor
\begin{equation}\label{Excitation F}
{\cal B}_{n} = \left(\frac{1}{2 p(\omega)} \frac{A^{(+)} (\omega)}{\frac{dA^{(-)} (\omega)}{d\omega}}   \right)_{\omega=\omega_n}
\end{equation}
but, in addition, it takes explicitly into account the role of the BH perturbation. We have
 \begin{equation}\label{EC}
{\cal C}_n={\cal B}_n \int_{-\infty}^{+\infty} \frac{\phi_0(r')\phi^\mathrm{in}_{\omega_n}(r')}{\sqrt{\omega_n/p(\omega_n)}A^{(+)}(\omega_n)}   dr'_\ast.
\end{equation}
For more precisions concerning the excitation factors (intrinsic quantities) and the excitation coefficients (extrinsic quantities), we refer to Refs.~\cite{Berti:2006wq,Decanini:2014bwa,Decanini:2014kha}.

\subsection{Numerical results and discussions}

\subsubsection{Numerical methods}
\label{Sec.IIB1}

To construct the waveform (\ref{reponse_partielle}), we have to obtain numerically the functions $\phi^\mathrm{in}_{\omega}(r)$ and $\phi^\mathrm{up}_{\omega}(r)$ as well as the coefficient $A^{(-)}(\omega)$ for $\omega \in \mathbb{R}^+$. This can be achieved by integrating numerically the Regge-Wheeler equation (\ref{RW}) with the Runge-Kutta method by using a sufficiently large working precision. It is necessary to initialize the process with Taylor series expansions converging near the horizon and to compare the solutions to asymptotic expansions with ingoing and outgoing behavior at spatial infinity. In order to obtain reliable results for ``large'' values of the mass parameter, it necessary to decode systematically, by Pad\'e summation, the information hidden in the divergent part of the asymptotic expansions considered but also to work very carefully for frequencies near the branch point $+ \mu$. Moreover, in Eq.~(\ref{reponse_partielle}), we have to discretize the integral over $\omega$. In order to obtain numerically stable waveforms, we can limit the range of frequencies to $-8\leq 2M\omega \leq+8$ and take for the frequency resolution $2M\delta\omega=1/10000$.

The quasinormal frequencies $\omega_{n}$ (as well as the complex frequencies of the QBSs) can be determined by using the method developed for massive fields by Konoplya and Zhidenko \cite{Konoplya:2004wg} and which can be numerically implemented by modifying the Hill determinant approach of Majumdar and Panchapakesan \cite{mp} (for more precision, see Sec.~II of Ref.~\cite{Decanini:2014bwa} as well as Appendixes B and C of Ref.~\cite{Decanini:2015yba}).

The coefficients $A^{(+)}(\omega_n) $, the excitation factors ${\cal B}_{n}$ and the excitation coefficients ${\cal C}_n$ can be obtained from $\phi^\mathrm{in}_{\omega}(r)$ by integrating numerically the Regge-Wheeler equation (\ref{RW}) for $\omega=\omega_{n}$ and $\omega=\omega_{n}+\epsilon$ (we have taken $\epsilon\sim 10^{-10}$) with the Runge-Kutta method and then by comparing the solution to asymptotic expansions (decoded by Pad\'e summation) with ingoing and outgoing behavior at spatial infinity.

To construct the ringing (\ref{TimeEvolution_QNM}), we need, in addition to the quasinormal frequencies $\omega_{n}$ and the excitation coefficients ${\cal C}_n$, the functions $\phi^\mathrm{up}_{\omega_n}(r)$. They can be obtained by noting that $\phi^\mathrm{up}_{\omega_n}(r) = \phi^\mathrm{in}_{\omega_n}(r)/A^{(+)}(\omega_n)$. It is also important to recall that the quasinormal contribution (\ref{TimeEvolution_QNM}) does not provide physically relevant results at ``early times'' due to its exponentially divergent behavior as $t$ decreases. In our previous works \cite{Decanini:2014bwa,Decanini:2014kha}, we have proposed to construct the starting time $t_\mathrm{start}$ of the BH ringing from the group velocity corresponding to the quasinormal frequency $\omega_{n}$ which is given by $v_\mathrm{g}=\mathrm{Re}[p(\omega_{n})]/\mathrm{Re}[\omega_{n}]$. By assuming again that the source is strongly localized while the observer is located at a rather large distance $r$ from the source, we can use for the starting time
\begin{equation}
\label{t_start}
 t_\mathrm{start} \approx \frac{r_\ast(r) + r_\ast(r_0)} {\mathrm{Re}[p(\omega_{n})]/\mathrm{Re}[\omega_{n}]}.
\end{equation}

\subsubsection{Numerical results and comments}

\begin{figure}[!h]
\includegraphics[scale=0.52]{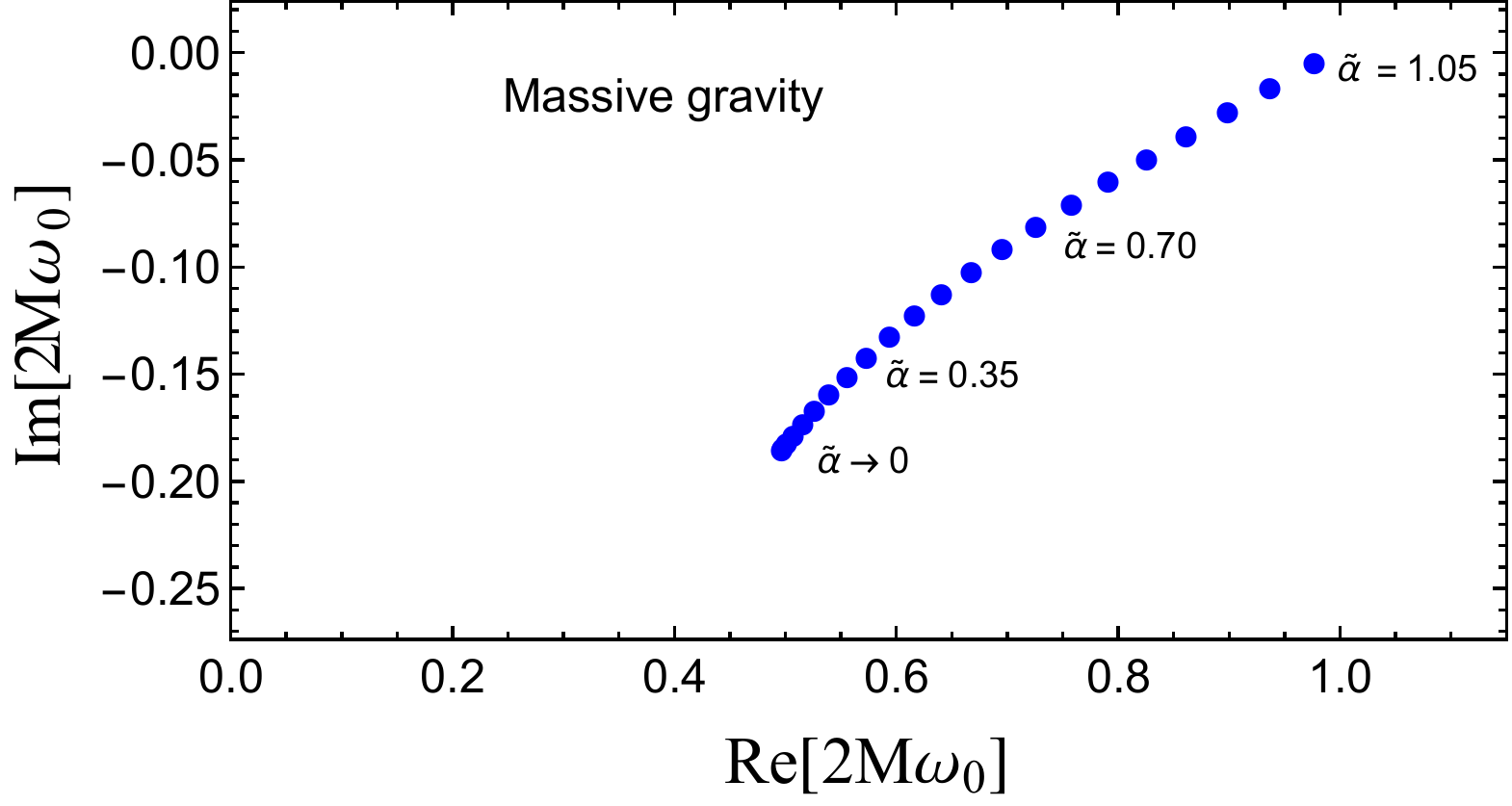}%[height=4.5cm,keepaspectratio=true]
\setlength\abovecaptionskip{0.25ex}
\vspace{-0.10cm}
\caption{\label{fig:OM_n=0} Complex frequency $\omega_0$ of the odd-parity $(\ell=1,n=0)$ QNM (massive spin-2 field). $2M\omega_0$ is followed from ${\tilde \alpha} \to 0$ to ${\tilde \alpha} =1.05$. Above ${\tilde \alpha} \approx 1.06$, the QNM disappears.}
\end{figure}

\begin{figure}[!t]
\centering
\includegraphics[scale=0.52]{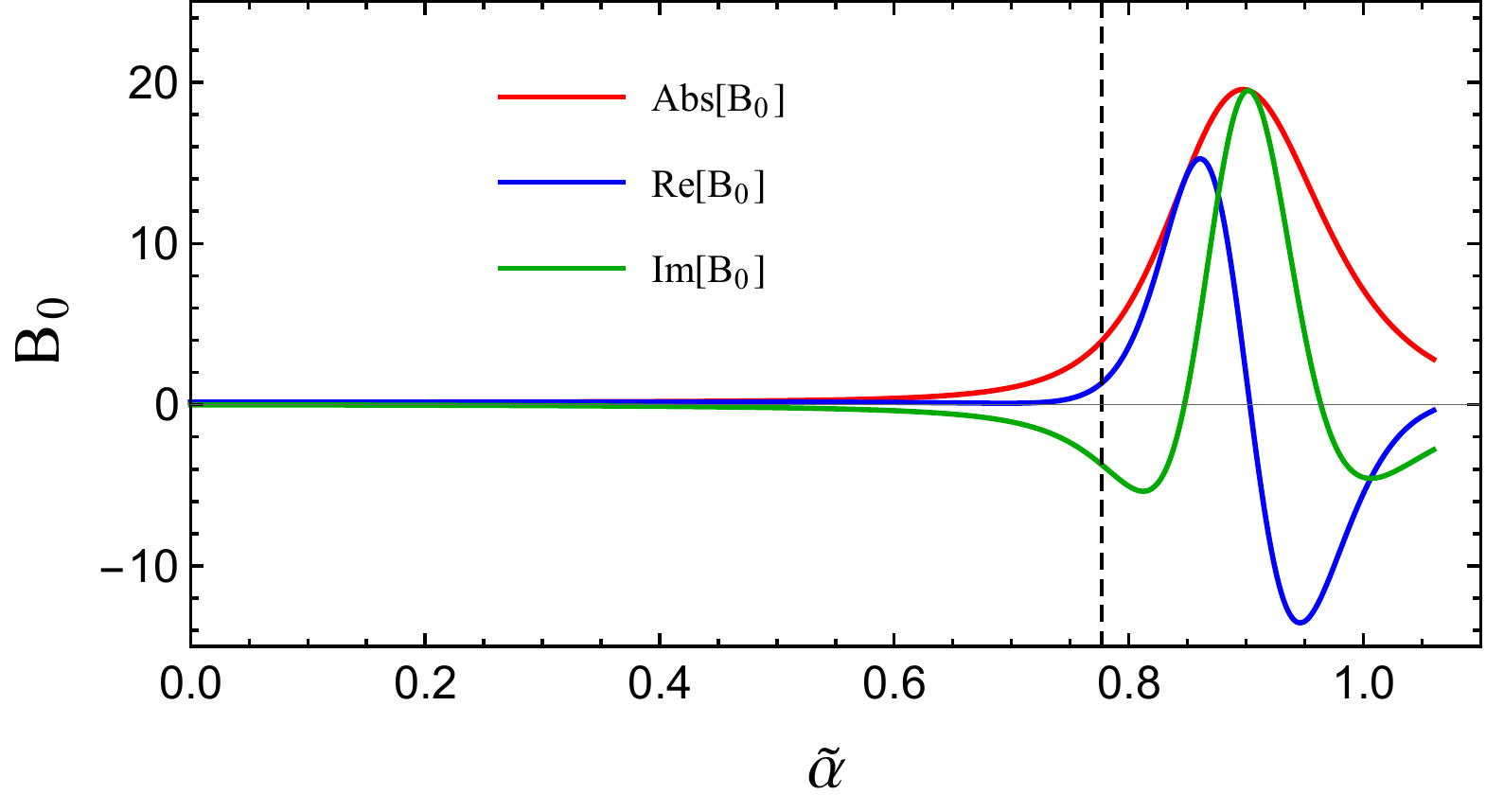}%[height=4.55cm,keepaspectratio=true]
\setlength\abovecaptionskip{0.25ex}
\vspace{-0.10cm}
\caption{\label{fig:B0_Exfact}  Resonant behavior, in massive gravity, of the excitation factor ${\cal B}_0$ of the odd-parity $(\ell=1,n=0)$ QNM. The maximum of $|2M{\cal B}_0|$ occurs for the critical value ${\tilde \alpha}_0 \approx 0.89757$; we then have $2M\omega_0 \approx 0.85969073 - 0.03878222 i$, $2M{\cal B}_0 \approx 3.25237 + 19.28190 i$ and $|2M{\cal B}_0| \approx 19.5543 $.}
\vspace{0.2cm}
\includegraphics[scale=0.52]{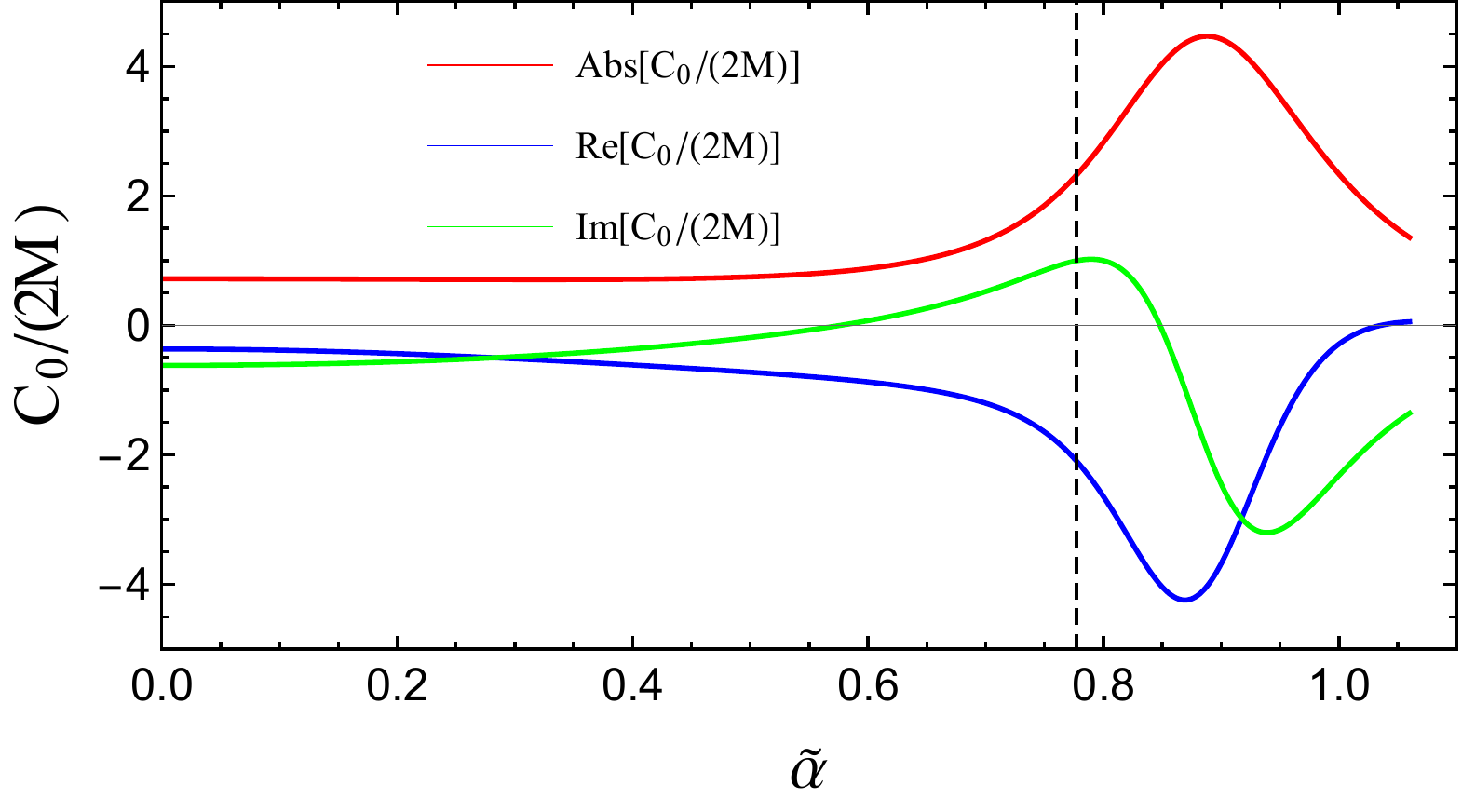}%[height=4.57cm,keepaspectratio=true]
\setlength\abovecaptionskip{0.25ex}
\vspace{-0.10cm}
\caption{\label{fig:C0_Excoeff} Resonant behavior, in massive gravity, of the excitation coefficient ${\cal C}_0$ of the odd-parity $(\ell=1,n=0)$ QNM. It is obtained from (\ref{EC}) by using (\ref{Cauchy_data}) with $\phi_0=1$, $a=1$ and $r_0=10M$. The maximum of $|2M{\cal C}_0|$ occurs for the critical value ${\tilde \alpha}_0 \approx 0.88808$; we then have $2M\omega_0 \approx 0.85277076 - 0.04084908 i$, $2M{\cal C}_0 \approx -4.02613 - 1.93037 i$ and $|2M{\cal C}_0| \approx 4.46498 $.}
\vspace{0.15cm}
\includegraphics[scale=0.52]{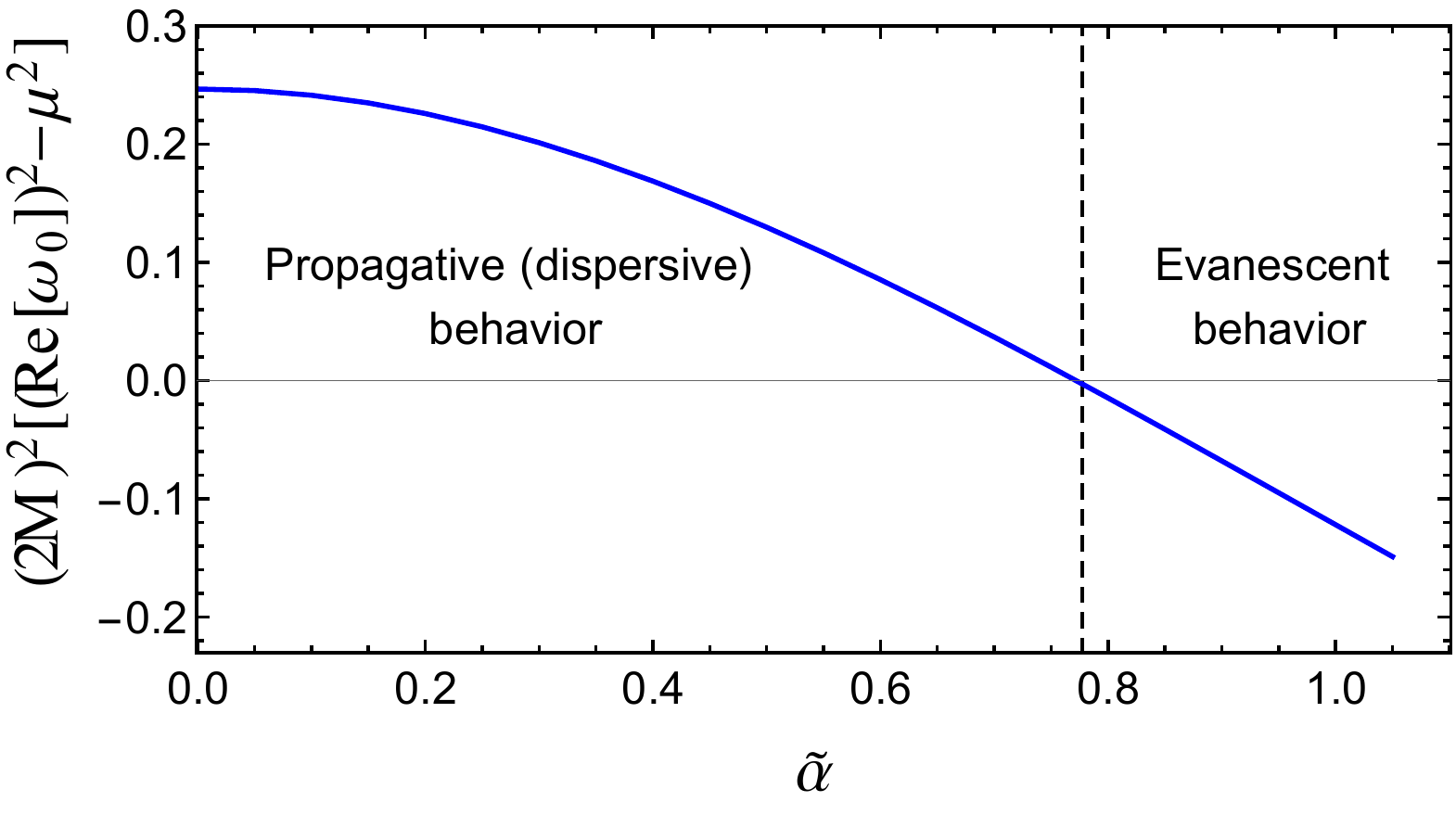}%[height=4.55cm,keepaspectratio=true]
\centering
\setlength\abovecaptionskip{0.25ex}
\vspace{-0.10cm}
\caption{\label{fig:ReQNM} The square of the wave number $p(\omega = \mathrm{Re}[\omega_0])$ as a function of the mass. In the low-mass regime, the partial wave exciting the quasinormal ringing has a propagative behavior while, for masses in the range where the excitation factor ${\cal B}_0$ and the excitation coefficient ${\cal C}_0$ have a strong resonant behavior, its has an evanescent behavior.}
\end{figure}

\begin{figure}[!t]
\centering
\includegraphics[scale=0.536]{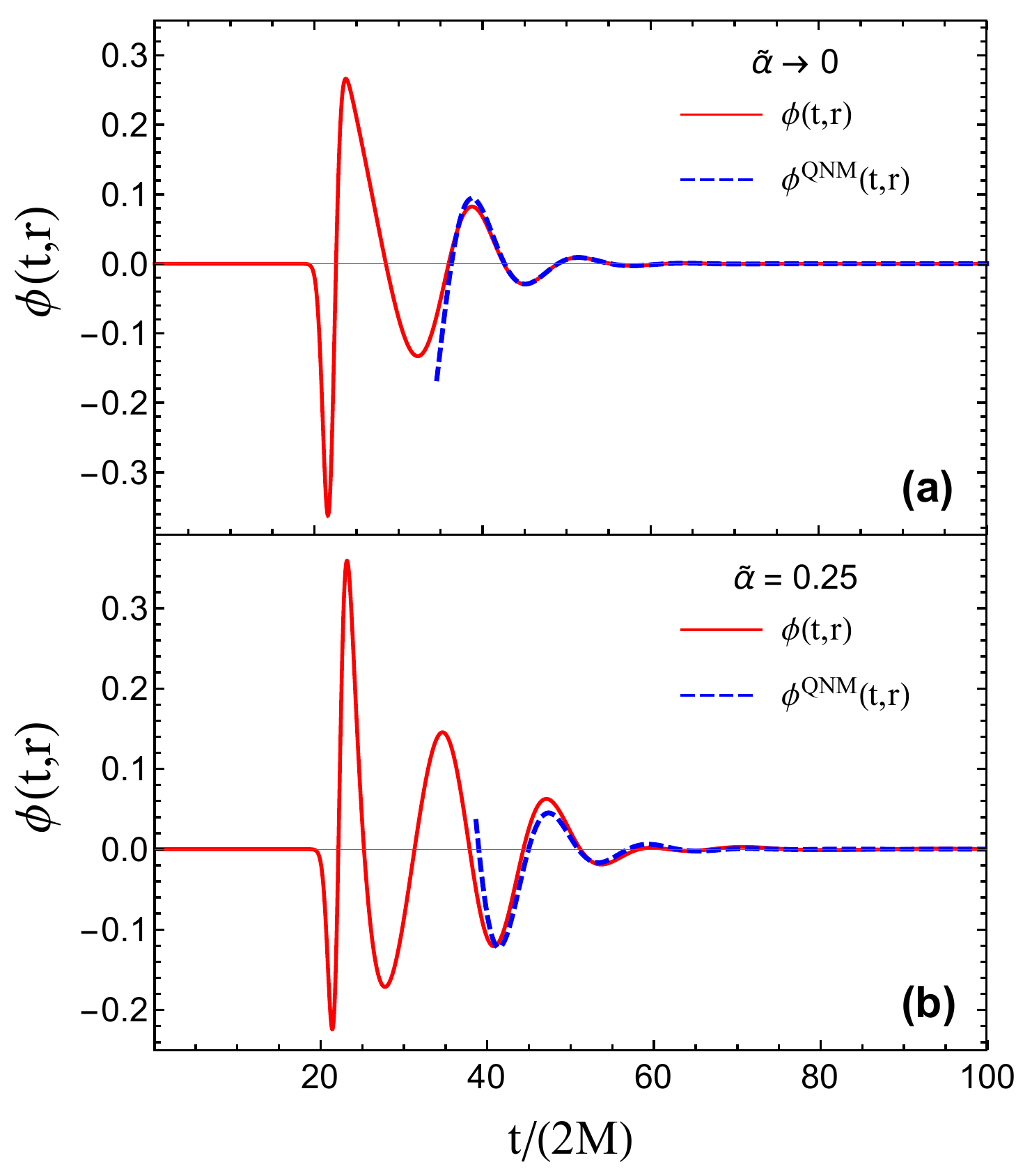}
\setlength\abovecaptionskip{0.25ex}
\vspace{-0.1cm}
\caption{\label{fig:Reponse_mu_00_025_QNM_Extrinsic}  Comparison of the waveform (\ref{reponse_partielle}) with the quasinormal waveform  (\ref{TimeEvolution_QNM}). The results are obtained for (a) $\tilde\alpha\rightarrow 0$ and (b) $\tilde\alpha = 0.25$. The parameters of the Gaussian source (\ref{Cauchy_data}) are $\phi_0=1$, $a=1$ and $r_0=10M$. The observer is located at $r=50M$. The quality of the superposition of the two signals decreases as the mass increases due to the dispersive nature of the massive field (the excitation of QBSs playing a negligible role).}
\vspace{-0.2cm}
\end{figure}

\begin{figure}[!t]
\centering
\includegraphics[scale=0.536]{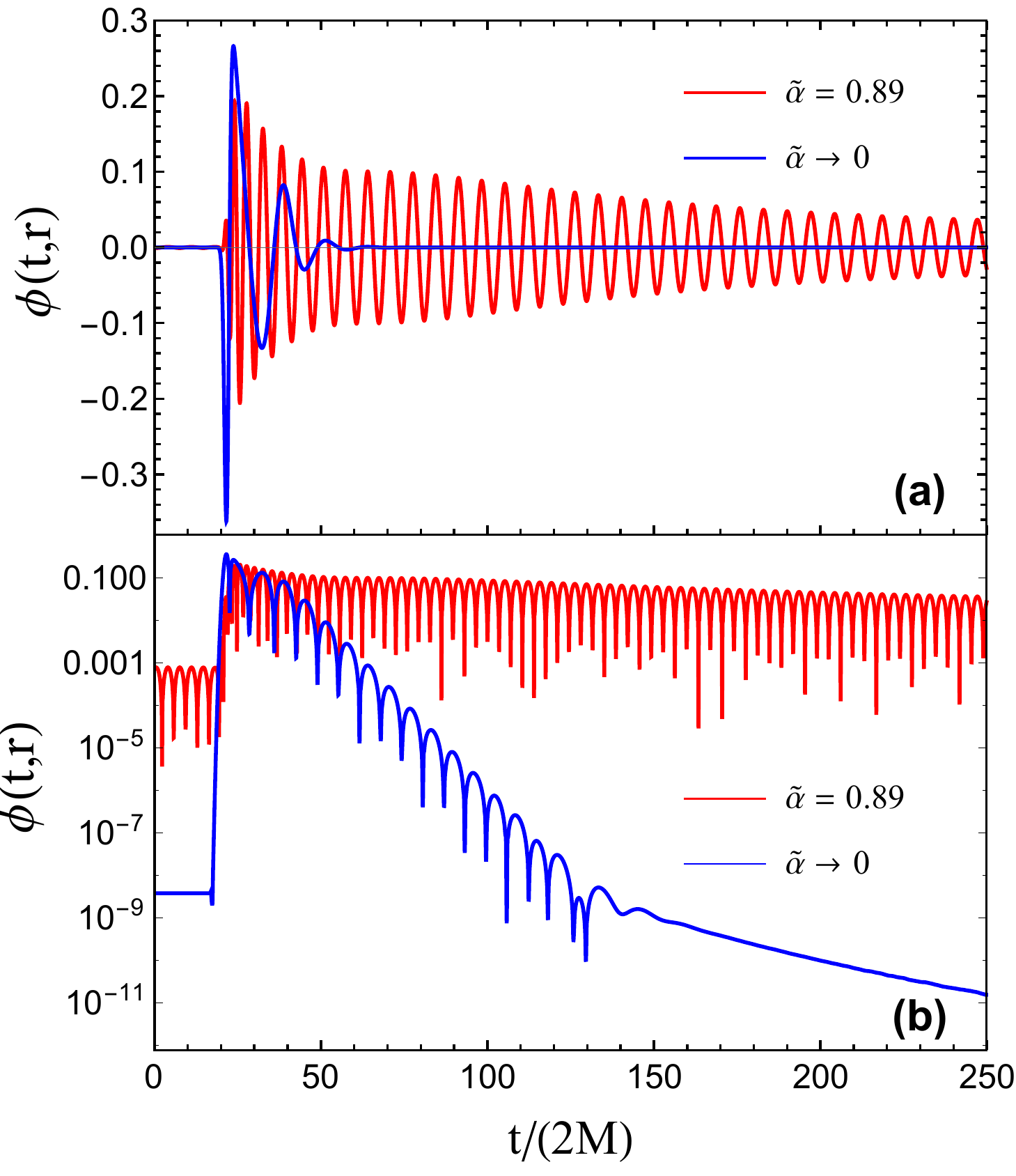}
\setlength\abovecaptionskip{0.25ex}
\vspace{-0.10cm}
\caption{\label{fig:Reponse_mu_0_089_log_Extrinsic} Comparison of the waveforms obtained for $\tilde{\alpha}\rightarrow0$ and for $\tilde{\alpha}=0.89$. The parameters of the Gaussian source (\ref{Cauchy_data}) are $\phi_0=1$, $a=1$ and $r_0=10M$. The observer is located at $r=50M$. (a) Normal plot and (b) semi-log plot.}
\vspace{-0.2cm}
\end{figure}

\begin{figure}[!t]
\centering
\vspace{0.15cm}
\includegraphics[scale=0.55]{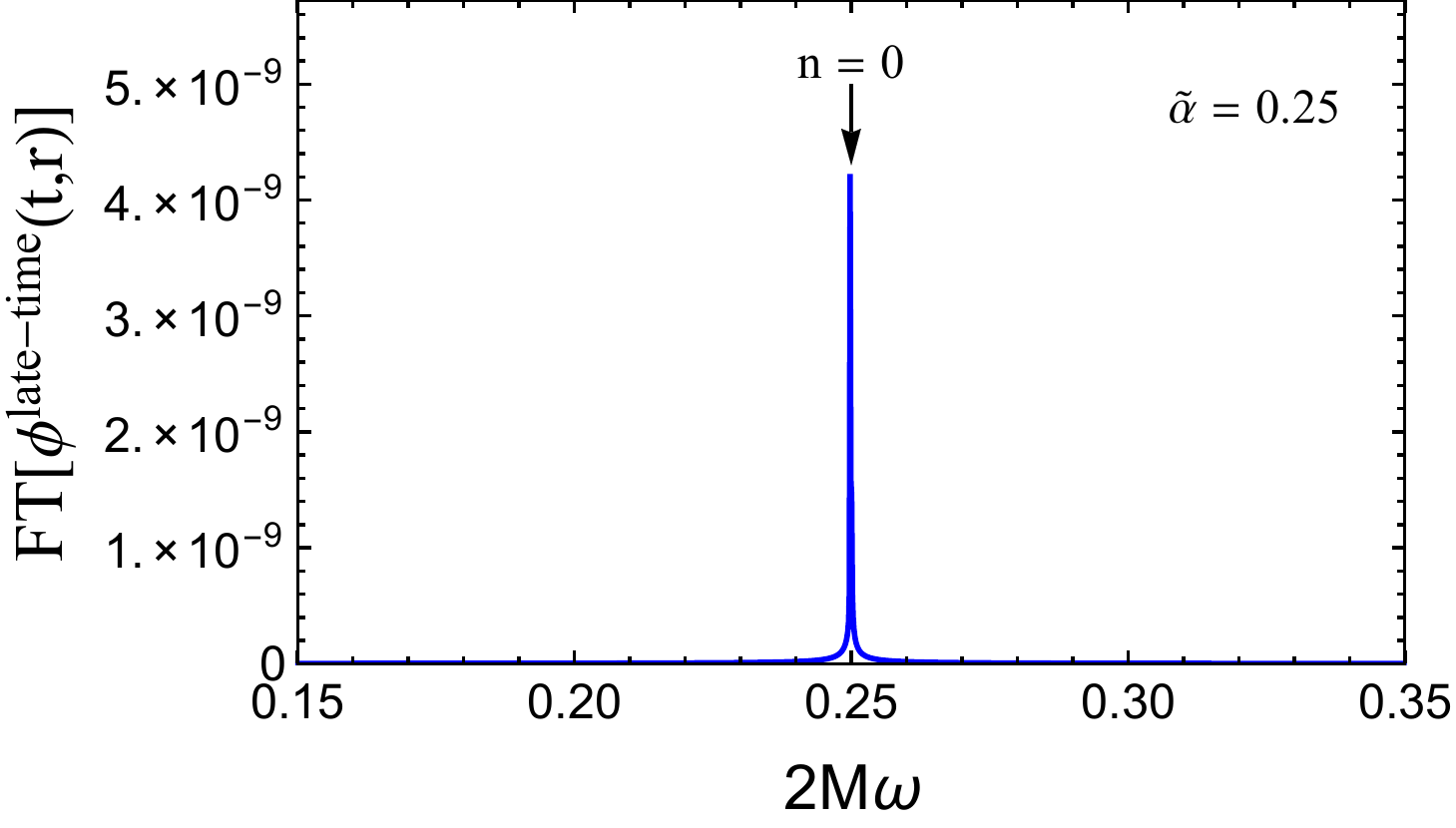}
\setlength\abovecaptionskip{0.25ex}
\vspace{-0.10cm}
\caption{\label{fig:FT_QBS_025}  The spectral content of the ``late-time'' phase of the waveform for $\tilde\alpha=0.25$. The parameters of the Gaussian source (\ref{Cauchy_data}) are $\phi_0=1$, $a=1$ and $r_0=10M$. The observer is located at $r=50M$. We only observe the signature of the first long-lived QBS (see Table~\ref{tab:table1}); it is weakly excited (note its very low amplitude) and has little influence on the waveform (see Fig.~\ref{fig:Reponse_mu_00_025_QNM_Extrinsic}).}
%\vspace{-0.2cm}
\end{figure}

\begin{figure}[!t]
\centering
\includegraphics[scale=0.50]{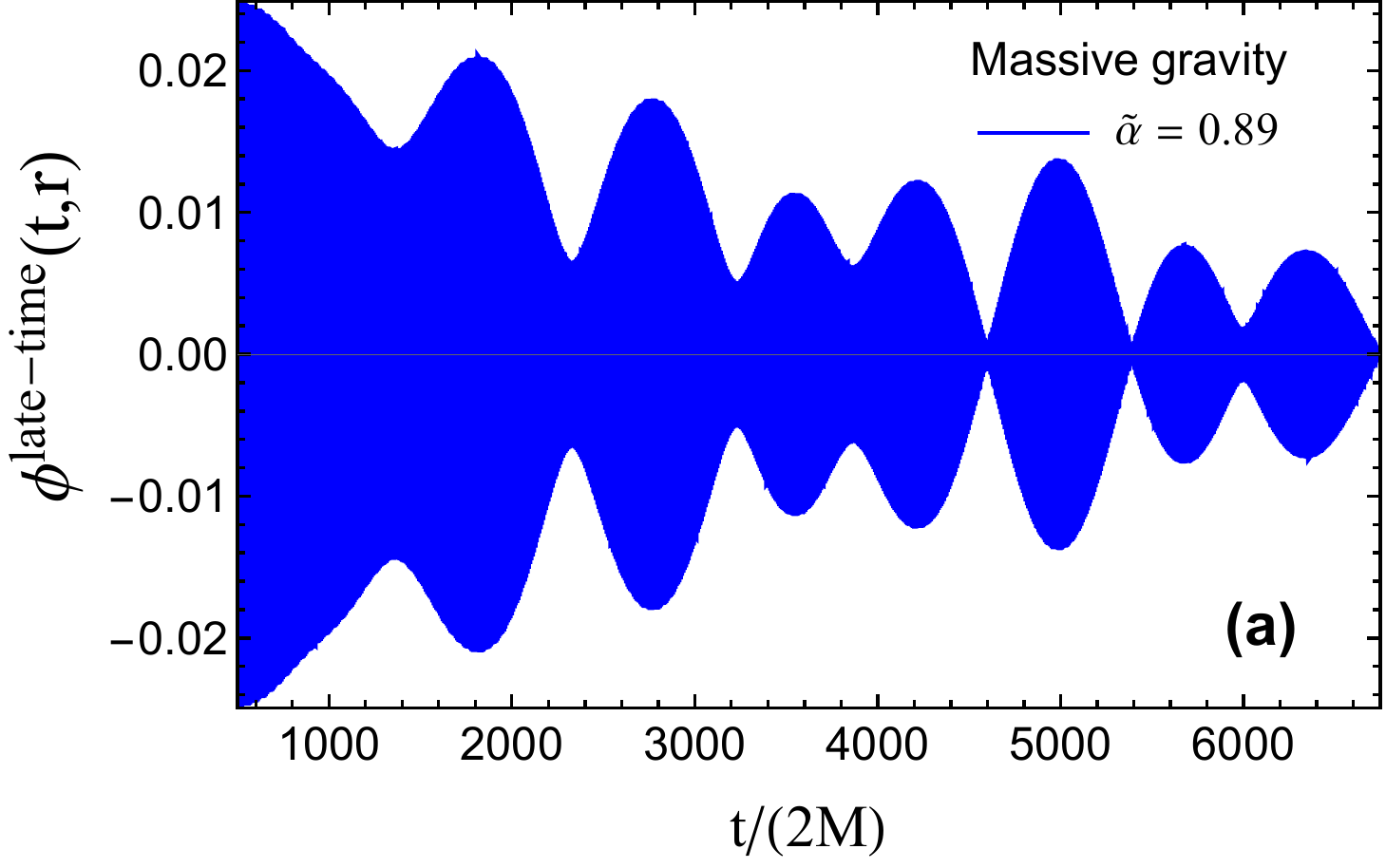}
\centering
\vspace{0.10cm}
\includegraphics[scale=0.55]{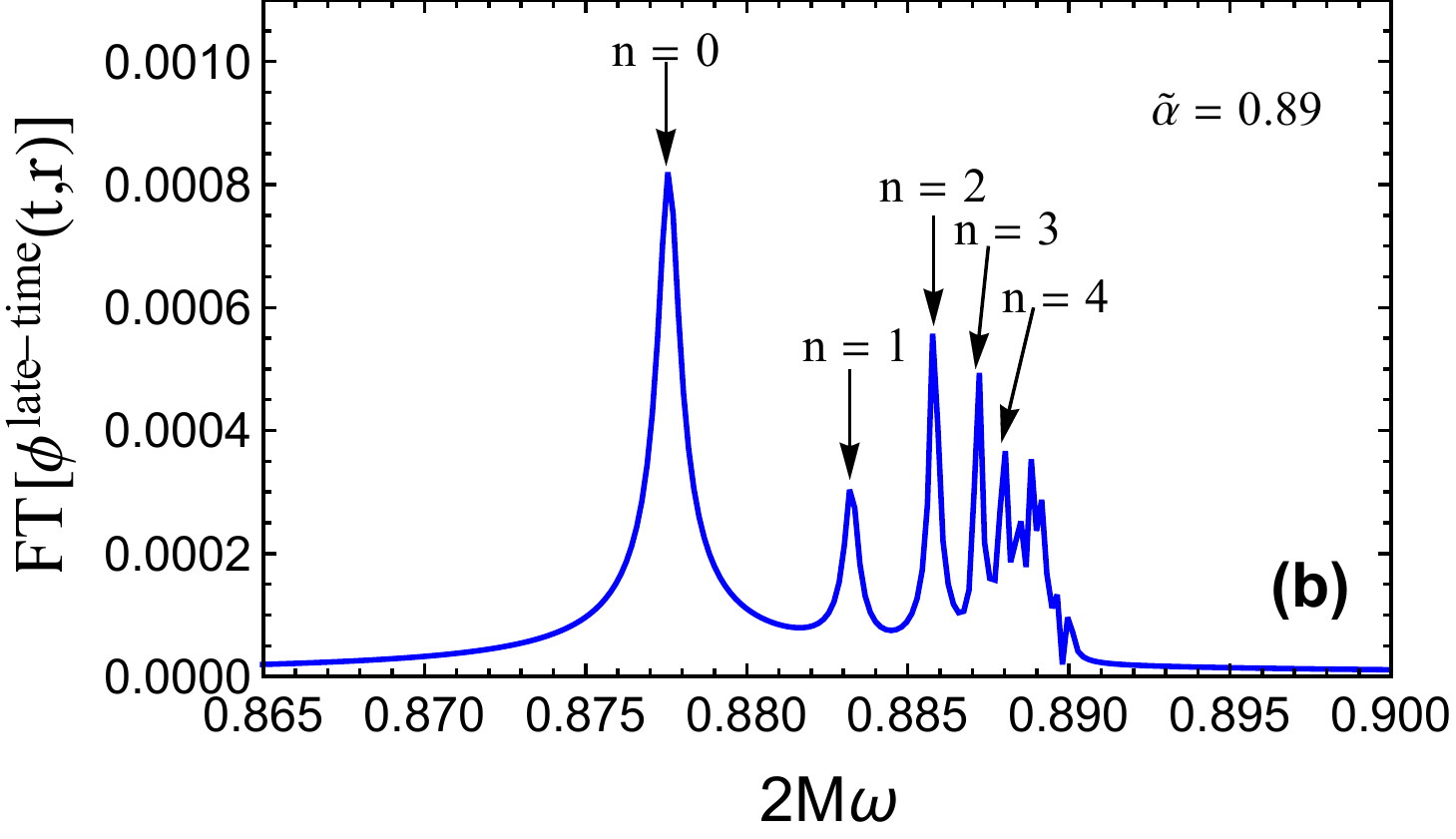}
\setlength\abovecaptionskip{0.25ex}
\vspace{-0.10cm}
\caption{\label{fig:FT_QBS_089}  (a) The late-time phase of the waveform for $\tilde\alpha=0.89$ and (b) the spectral content of the full waveform. The parameters of the Gaussian source (\ref{Cauchy_data}) are $\phi_0=1$, $a=1$ and $r_0=10M$. The observer is located at $r=50M$. We observe the signature of the first long-lived QBSs (see also Table~\ref{tab:table1}) and beats due to interference between QBSs of neighboring frequencies.}
\end{figure}

In Fig.~\ref{fig:OM_n=0}, we display the effect of the graviton mass on the complex frequency $\omega_0$ of the fundamental QNM and in Fig.~\ref{fig:B0_Exfact}, we exhibit the strong resonant behavior of the associated excitation factor ${\cal B}_0$ occurring around the critical value ${\tilde \alpha_0} \approx 0.90$. Here, we focus on the least damped QNM but it is worth noting that the same kind of quasinormal resonant behavior also exists for the overtones but with excitation factors ${\cal B}_n $ of much lower amplitude. In Fig.~\ref{fig:C0_Excoeff}, we exhibit the strong resonant behavior of the excitation coefficient ${\cal C}_0$ for particular values of the parameters defining the initial data (\ref{Cauchy_data}). It occurs around the critical value ${\tilde \alpha_0} \approx 0.89$ and is rather similar to the behavior of the corresponding excitation factor ${\cal B}_0$. It depends very little on the parameters defining the Cauchy problem. Of course, for overtones, the quasinormal resonant behavior is more and more attenuated as the overtone index $n$ increases. It is also important to note that the quasinormal resonant behavior occurs for masses in a range where the fundamental QNM is a long-lived mode (see Fig.~\ref{fig:OM_n=0}). From a theoretical point of view, if we focus our attention exclusively on Eq.~(\ref{TimeEvolution_QNM}) (see also Refs.~\cite{Decanini:2014bwa,Decanini:2014kha}), it is logical to think that this leads to giant and slowly decaying ringings. In fact, this way of thinking is rather naive and it seems that, in waveforms, it is not possible to exhibit such extraordinary ringings for two main reasons (here we restrict our discussion to the fundamental QNM because it provides the most interesting contribution):

\begin{enumerate}[label=(\roman*)]

\item The quasinormal ringing (\ref{TimeEvolution_QNM}) is excited when a real frequency $\omega$ in the integral (\ref{reponse_partielle}) defining the waveform coincides with (or is very close to) the excitation frequency $\mathrm{Re}[\omega_0]$ of the $n=0$ QNM. In the low-mass regime, the wave number $p(\omega = \mathrm{Re}[\omega_0])$ is a real positive number and the partial wave which excites the ringing has a propagative behavior (see Fig.~\ref{fig:ReQNM}). The ringing can be clearly identify in the waveform (see Fig.~\ref{fig:Reponse_mu_00_025_QNM_Extrinsic}) even if, as the mass parameter increases, the quality of the superposition of the signals decreases. For masses in the range where the excitation factor ${\cal B}_0$ and the excitation coefficient ${\cal C}_0$ have a strong resonant behavior, the wave number $p(\omega = \mathrm{Re}[\omega_0])$ is an imaginary number (the real part of the quasinormal frequency is smaller than the mass parameter and lies into the cut of the retarded Green function) and, as a consequence, the partial wave which could excite the ringing has an evanescent behavior (see Fig.~\ref{fig:ReQNM} as well as Figs.~\ref{fig:B0_Exfact} and \ref{fig:C0_Excoeff}). Theoretically, this leads to a significant attenuation of the ringing amplitude in the waveform. In Fig.~\ref{fig:Reponse_mu_0_089_log_Extrinsic}, we display the waveform for a value of the reduced mass ${\tilde \alpha}$ very close to the critical value ${\tilde \alpha}_0$. We cannot identify the ringing but we can, however, observe that the amplitude of the waveform is more larger than in the massless limit and that it decays very slowly. Such a behavior is a consequence of the excitation of QBSs (see below).
    
\item For any nonvanishing value of the reduced mass ${\tilde \alpha}$, the QBSs of the Schwarzschild BH are excited. Of course, their influence is negligible for ${\tilde \alpha} \to 0$ (see Table~\ref{tab:table1} and Fig.~\ref{fig:Reponse_mu_00_025_QNM_Extrinsic}) but increases with ${\tilde \alpha}$ (see Fig.~\ref{fig:FT_QBS_025} where we displays the spectral content of the late-time tail of the waveform for $\tilde\alpha=0.25$) and, for higher values of ${\tilde \alpha}$, they can even blur the QNM contribution (as we have already noted in another context in Ref.~\cite{Decanini:2015yba}). But near and above the critical value ${\tilde \alpha}_0$ of the reduced mass, the QBSs of the BH not only blur the QNM contribution but provide the main contribution to waveforms (see Figs.~\ref{fig:Reponse_mu_0_089_log_Extrinsic} and \ref{fig:FT_QBS_089}).
    
\end{enumerate}

\begin{figure}[!t]
\centering
\includegraphics[scale=0.536]{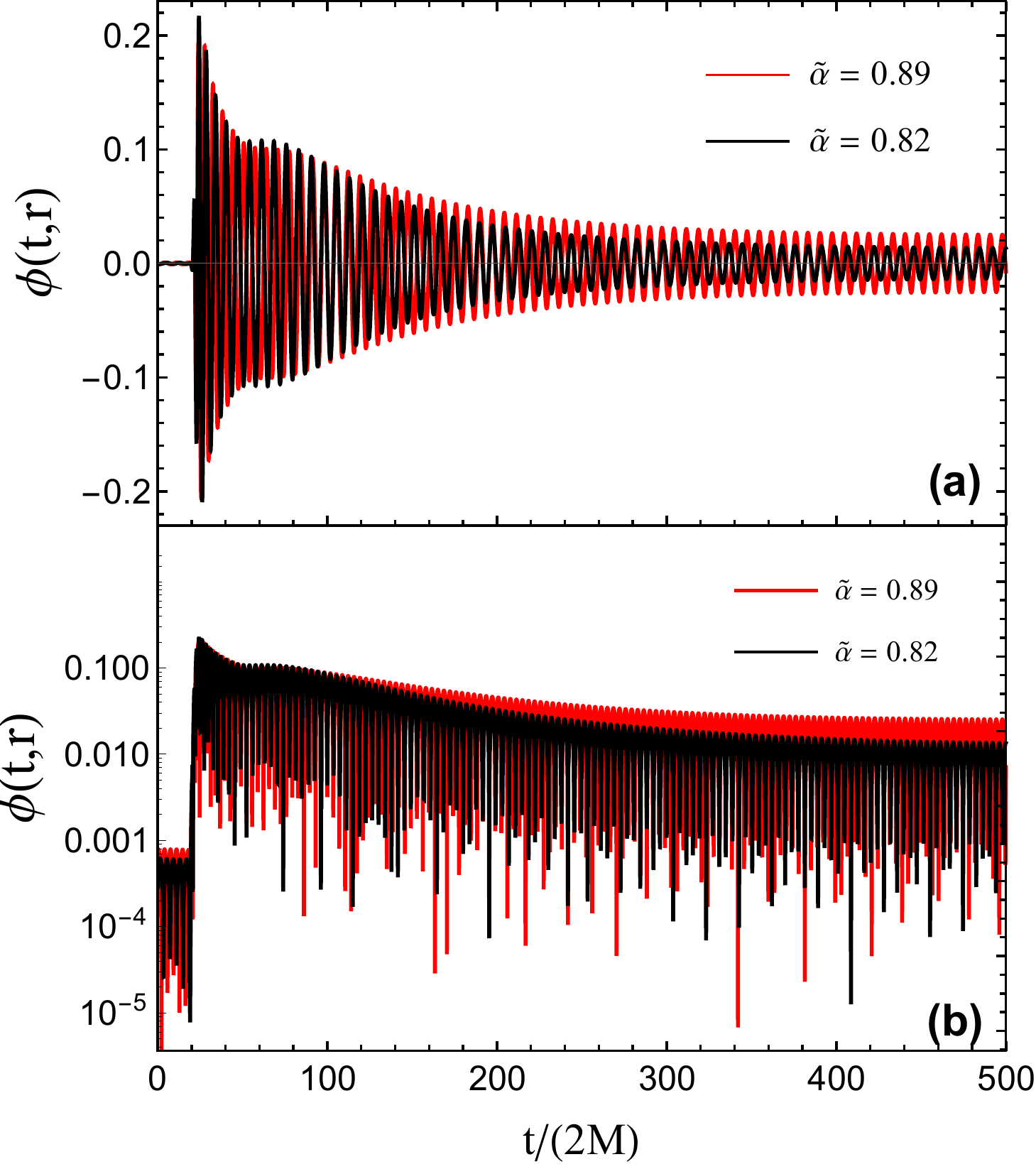}
\setlength\abovecaptionskip{0.25ex}
\vspace{-0.10cm}
\caption{\label{fig:Reponse_mu_082_089_log_Extrinsic}  Comparison of the waveforms obtained for $\tilde{\alpha}=0.89$ and for $\tilde{\alpha}=0.82$.
 The parameters of the Gaussian source (\ref{Cauchy_data}) are $\phi_0=1$, $a=1$ and $r_0=10M$. The observer is located at $r=50M$. (a) Normal plot and (b) semi-log plot.}
\end{figure}

\begin{figure}[!t]
\centering
\includegraphics[scale=0.55]{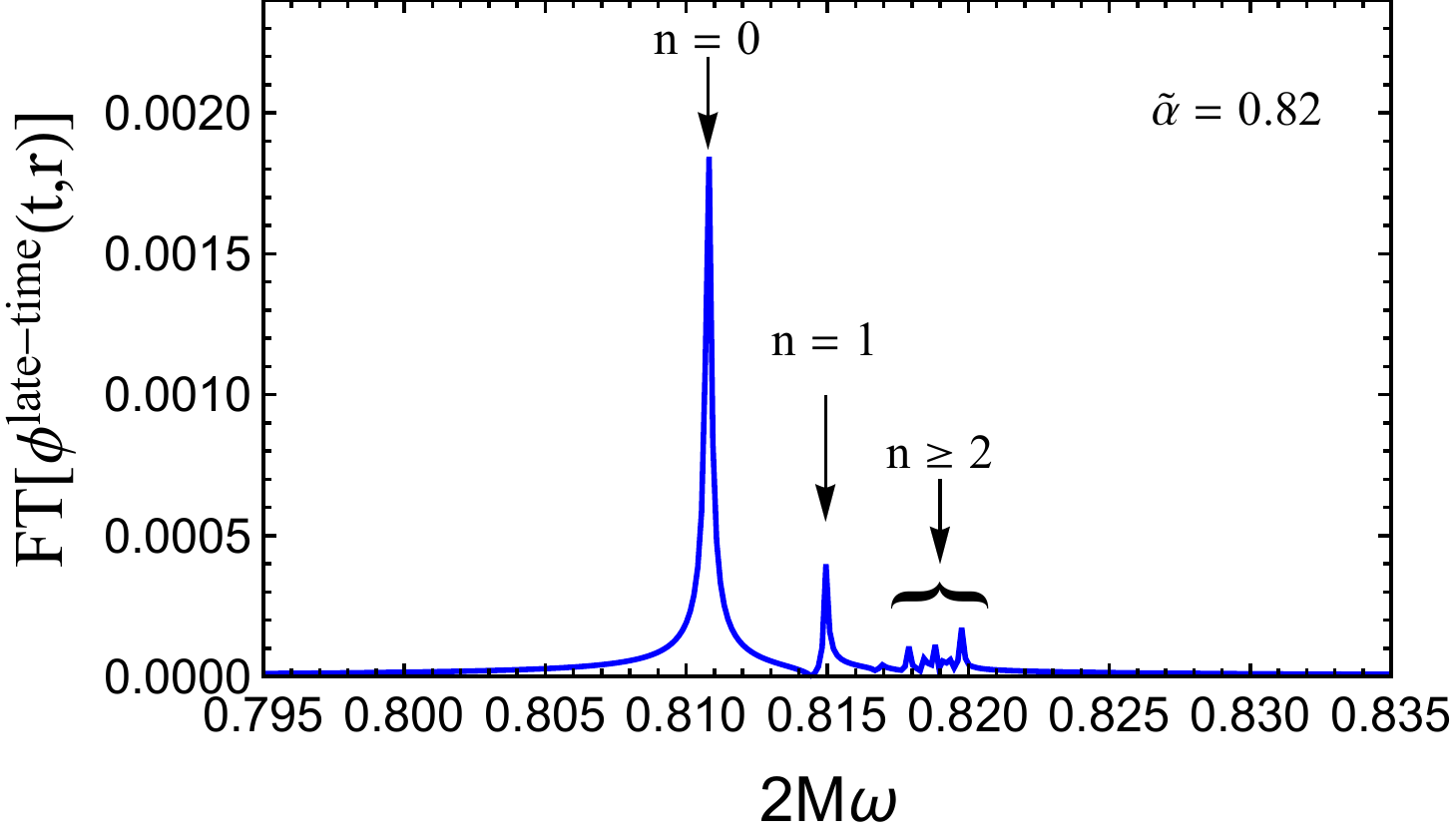}
\setlength\abovecaptionskip{0.25ex}
\vspace{-0.10cm}
\caption{\label{fig:FT_QBS_082}  The spectral content of the full waveform for $\tilde\alpha=0.82$ (see also Table~\ref{tab:table1}). The parameters of the Gaussian source (\ref{Cauchy_data}) are $\phi_0=1$, $a=1$ and $r_0=10M$. The observer is located at $r=50M$. We observe, in particular, that the first long-lived QBSs are not excited.}
\vspace{-0.3cm}
\end{figure}

\begin{figure}[!t]
\centering
\includegraphics[scale=0.536]{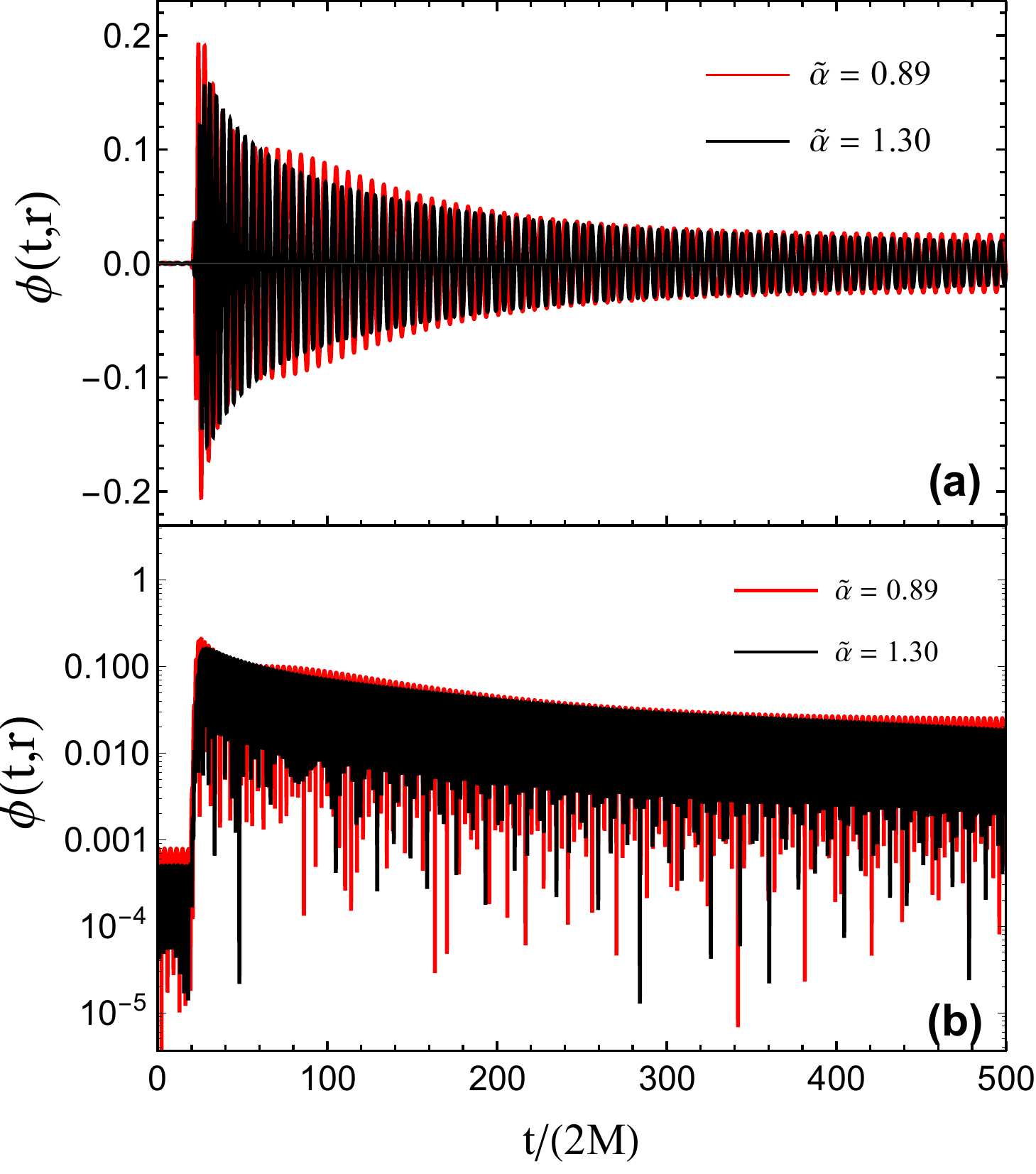}
\setlength\abovecaptionskip{0.25ex}
\vspace{-0.10cm}
\caption{\label{fig:Reponse_mu_130_089_log_Extrinsic}  Comparison of the waveforms obtained for $\tilde{\alpha}=0.89$ and for $\tilde{\alpha}=1.30$.
 The parameters of the Gaussian source (\ref{Cauchy_data}) are $\phi_0=1$, $a=1$ and $r_0=10M$. The observer is located at $r=50M$. (a) Normal plot and (b) semi-log plot.}
\end{figure}

\begin{figure}[!t]
\centering
\includegraphics[scale=0.55]{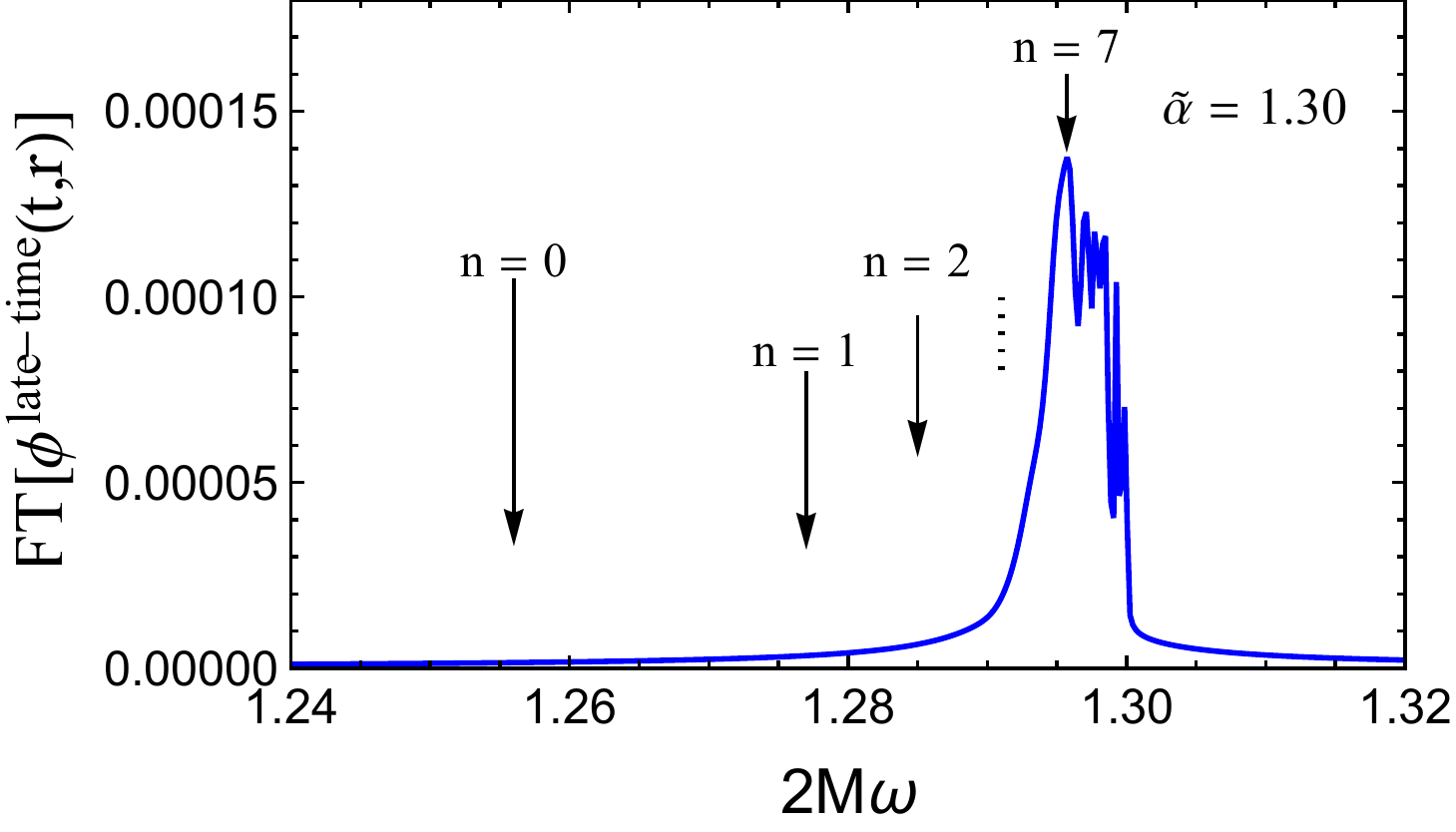}
\setlength\abovecaptionskip{0.25ex}
\vspace{-0.10cm}
\caption{\label{fig:FT_QBS_130}  The spectral content of the full waveform for $\tilde\alpha=1.30$ (see also Table~\ref{tab:table1}). The parameters of the Gaussian source (\ref{Cauchy_data}) are $\phi_0=1$, $a=1$ and $r_0=10M$. The observer is located at $r=50M$. We observe, in particular, that the first long-lived QBSs are not excited.}
\vspace{-0.2cm}
\end{figure}

It is interesting to also consider waveforms for reduced mass parameters:

\begin{enumerate}[label=(\roman*)]

  \item Near the critical value ${\tilde\alpha}_0$ but outside the stability domain (see Figs.~\ref{fig:Reponse_mu_082_089_log_Extrinsic} and \ref{fig:FT_QBS_082} where  we display the waveform corresponding to ${\tilde \alpha} = 0.82$ and its spectral content).
      
  \item Far above the critical value ${\tilde \alpha}_0$  (see Figs.~\ref{fig:Reponse_mu_130_089_log_Extrinsic} and \ref{fig:FT_QBS_130} where  we display the waveform corresponding to ${\tilde \alpha} = 1.30$ and its spectral content) and, in particular, for values for which the fundamental QNM does not exist (see Fig.~\ref{fig:OM_n=0}).
      
\end{enumerate}

\noindent In both cases, we can observe the neutralization of the giant ringing. It is worth noting that the amplitude of the waveforms is smaller than that corresponding to the critical value ${\tilde \alpha}_0$. In fact, we can observe that this amplitude increases from ${\tilde \alpha} \to 0$ to ${\tilde \alpha} \approx {\tilde \alpha}_0$ and then decreases from ${\tilde \alpha} = {\tilde \alpha}_0$ to ${\tilde \alpha} \to \infty$. It reaches a maximum for the critical mass parameter ${\tilde \alpha}_0$. In our opinion, this fact is reminiscent of the theoretical existence of giant ringings. We can also observe in Fig.~\ref{fig:FT_QBS_130} that the first long-lived QBSs are not excited. Indeed, they disappear because (i) their complex frequencies lie deeper in the complex plane and (ii) the real part of their complex frequencies is more smaller than the mass parameter and lies into the cut of the retarded Green function (see Table~\ref{tab:table1}). As a consequence, the partial waves which could excite them have an evanescent behavior. It is the mechanism which operates for the fundamental QNM around the critical value ${\tilde \alpha}_0$ and which leads to the nonobservability of giant ringings.

\begin{table}[!h]
\caption{\label{tab:table1}  Odd-parity $\ell=1$ mode of massive gravity. A sample of the first quasibound frequencies $\omega_{\ell n}$.}
\smallskip
\centering
\begin{tabular}{lcccccr}
\hline
\hline
& $(\ell,n)$  &$ {\tilde \alpha}$ & $2M \omega_{\ell n}$  &\\
\hline
& $(1,n)$ & $0$  & $ / $  &\\
\hline
& $(1,0)$ & $0.25$   & $0.24978\, - 9.37148\times 10^{-13} i$  &\\
& $(1,1)$ &              & $0.24988\, - 5.63842\times 10^{-13} i$  &\\
& $(1,2)$ &              & $0.24992\, - 3.30927\times 10^{-13} i$ &\\
& $(1,3)$ &              & $0.24995\, - 2.05049\times 10^{-13} i$ &\\
& $(1,4)$ &              & $0.24996\, - 1.34298\times 10^{-13} i$ &\\
\hline
& $(1,0)$ & $0.82$  &  $0.81077\, - 0.00007 i$  &\\
& $(1,1)$ &              & $0.81494\, - 0.00004 i$  &\\
& $(1,2)$ &              & $0.81684\, - 0.00003 i$  &\\
& $(1,3)$ &              & $0.81784\,- 0.00002 i$  &\\
& $(1,4)$ &              & $0.81844\, - 0.00001i$  &\\
\hline
& $(1,0)$ & $0.89$  & $0.87756\, - 0.00030 i$  &\\
& $(1,1)$ &              & $0.88324\, - 0.00019 i$  &\\
& $(1,2)$ &              & $0.88580\, - 0.00011 i$  &\\
& $(1,3)$ &              & $0.88715\, - 0.00006 i$  &\\
& $(1,4)$ &              & $0.88795\, - 0.00004 i$  &\\
\hline
& $(1,0)$ & $1.30$  & $1.25689\, - 0.01719 i$  &\\
& $(1,1)$ &              & $1.27712\, - 0.00724 i$  &\\
& $(1,2)$ &              & $1.28590\, - 0.00362 i $ &\\
& $(1,3)$ &              & $1.29049\, - 0.00204 i $ &\\
& $(1,4)$ &              & $1.29317\, - 0.00125 i$ &\\
\hline
\hline
\end{tabular}%
\end{table}

\section{Conclusion}

\begin{figure}[!h]
\centering
\includegraphics[scale=0.52]{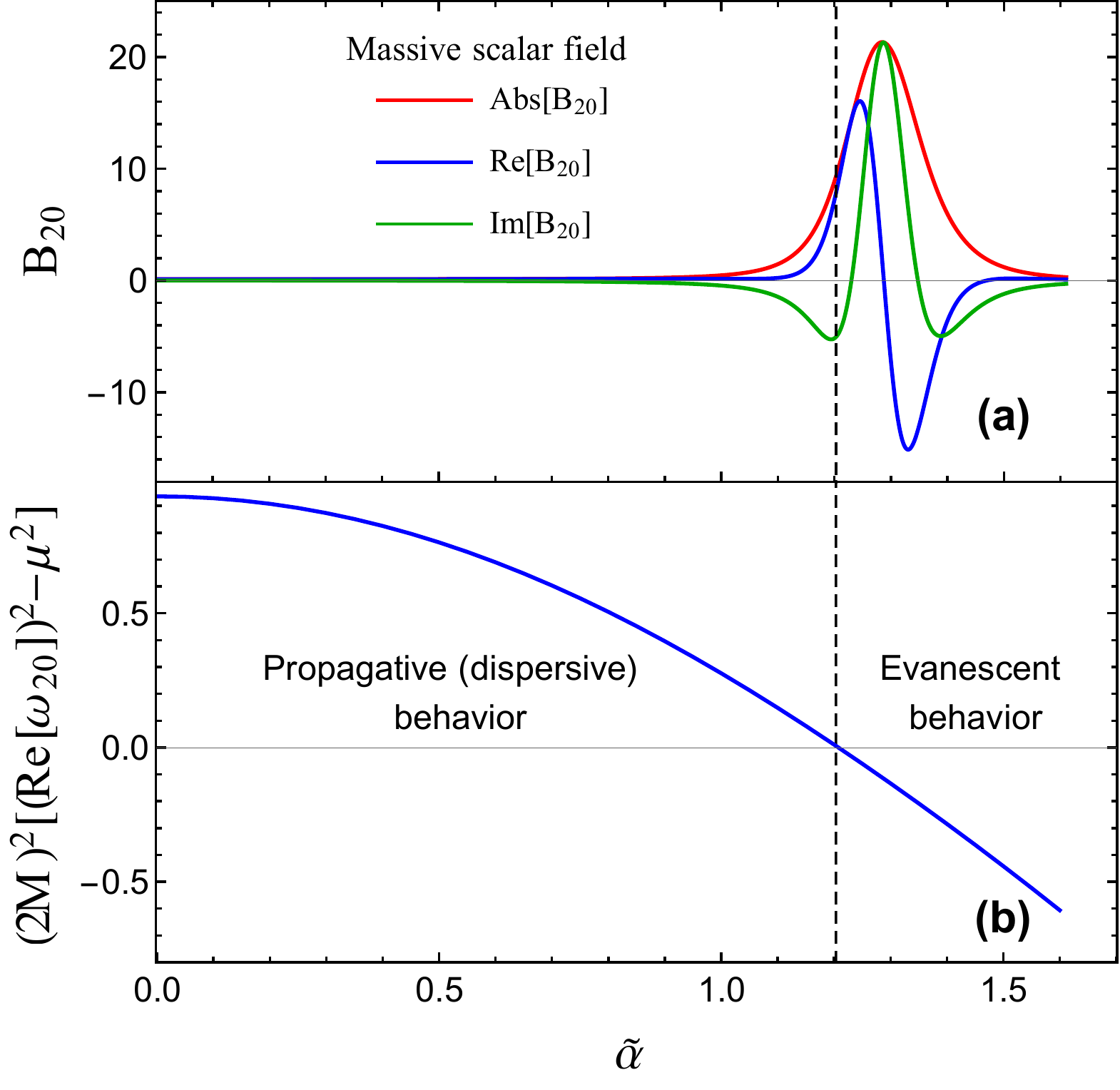}
\setlength\abovecaptionskip{0.25ex}
\vspace{-0.10cm}
\caption{\label{fig:B_Remu_20} The $(\ell=2,n=0)$ QNM of the massive scalar field. We denote by $\omega_{20}$ its complex frequency and by ${\cal B}_{20}$ the associated excitation factor. (a) Resonant behavior of ${\cal B}_{20}$. (b) The square of the wave number $p(\omega = \mathrm{Re}[\omega_{20}])$ as a function of the mass parameter. For masses in the range where the excitation factor ${\cal B}_{20}$ has a strong resonant behavior, the partial wave exciting the quasinormal ringing has an evanescent behavior. This leads to a significant attenuation of the ringing amplitude in the waveform (see Fig.~\ref{fig:Reponse22mu128_FT_Adiab_Tail}).}
\end{figure}

In this article, we have shown that the giant and slowly decaying ringings which could be generated in massive gravity due to the resonant behavior of the quasinormal excitation factors of the Schwarzschild BH are neutralized in waveforms. This is mainly a consequence of the coexistence of two effects which occur in the frequency range of interest: (i) the excitation of the QBSs of the BH and (ii) the evanescent nature of the particular partial modes which could excite the concerned QNMs. It should be noted that this neutralization process occurs for values of the reduced mass parameter $\tilde\alpha$ into the BH stability range (we have considered $\tilde\alpha=0.89$ and $\tilde\alpha=1.30$) but also outside this range (we have considered $\tilde\alpha=0.82$). Despite the neutralization, the waveform characteristics remain interesting from the observational point of view.

It is also interesting to note that, for values of $\tilde\alpha$ below and much below the threshold value ${\tilde \alpha}_t$ (we have considered $\tilde\alpha=0.25$ and $\tilde\alpha \to 0$ corresponding to the weak instability regime for the BH), the situation is very different. Of course, the ringing is neither giant nor slowly decaying but it is not blurred by the QBS contribution. As a consequence, it could be clearly observed in waveforms and used to test massive gravity theories with gravitational waves even if the graviton mass is very small.

\begin{figure*}[!t]
\includegraphics[scale=0.55]{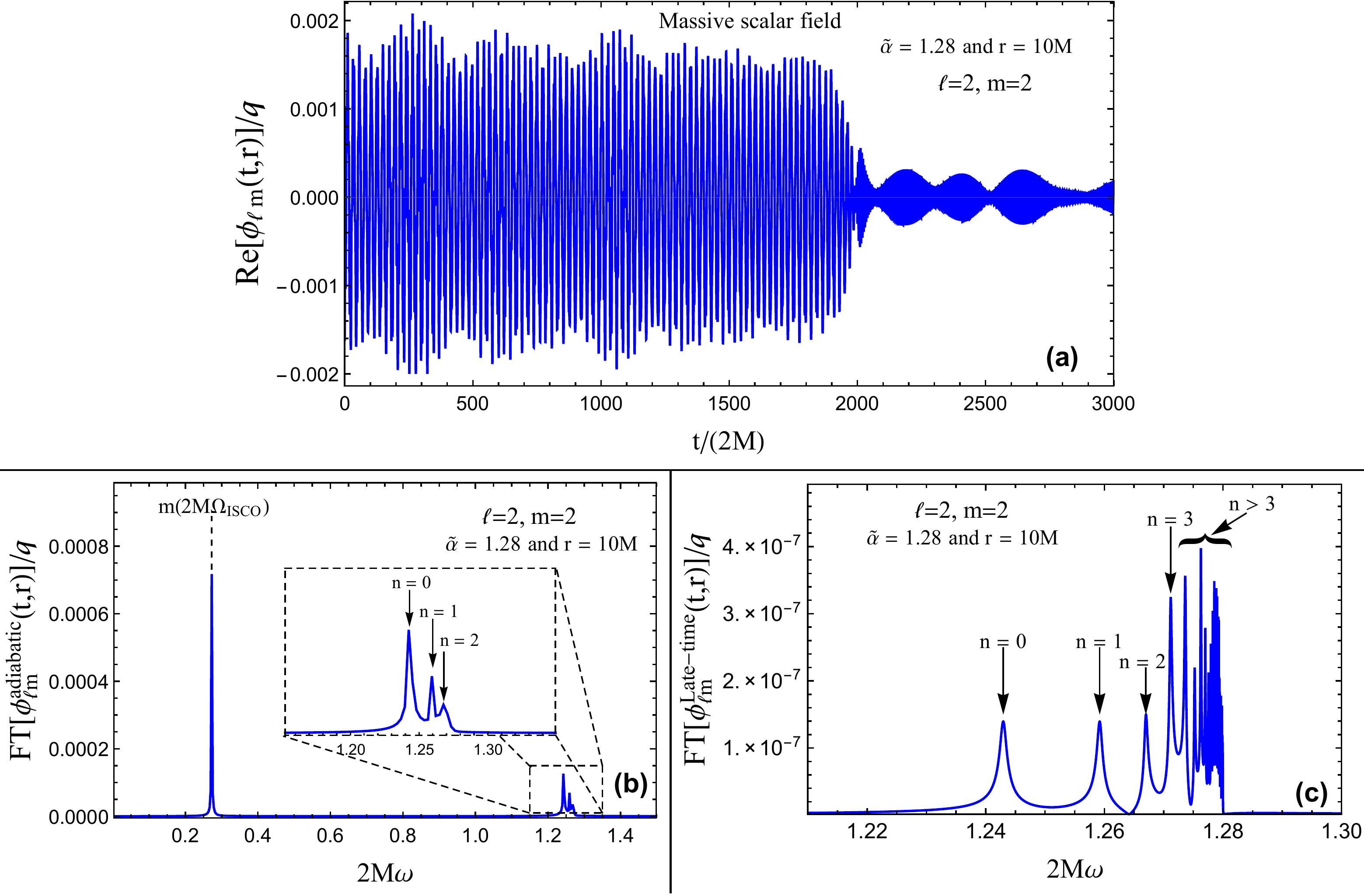}
\setlength\abovecaptionskip{0.25ex}
\vspace{-0.1cm}
\caption{\label{fig:Reponse22mu128_FT_Adiab_Tail} Quadrupolar waveform $\phi_{22}(t,r)$ associated with the $(\ell=2,m=2)$
mode of the massive scalar field and generated by a scalar point particle on a plunge trajectory (see Ref.~\cite{Decanini:2015yba} for the theory). The mass parameter corresponds to the maximum
of $|{\cal B}_{20}|$ (see Fig.~\ref{fig:B_Remu_20}) and the observer is located at $r=10M$. (a) The quasinormal ringing does not appear in the waveform. The beats are caused
 by interferences between QBSs. (b) Spectral content of the adiabatic phase. We observe, in addition to the signature of the quasicircular motion of the plunging particle,
 that of the first long-lived QBSs. (c) Spectral content of the late-time phase. We observe a profusion of long-lived QBSs with an accumulation which converges to
  the limiting frequency $2M\omega = \tilde\alpha$.}
\end{figure*}

In order to simplify our task, we have restricted our study to the odd-parity $\ell=1$ partial mode of the Fierz-Pauli theory in the Schwarzschild spacetime (here it is important to recall that its behavior is governed by a single differential equation of the Regge-Wheeler type [see Eq.~(\ref{Phi_ell1})] while all the other partial modes are governed by two or three coupled differential equations depending on the parity sector and the angular momentum) and we have, moreover, described the distortion of the Schwarzschild BH by an initial value problem. Of course, it would be very interesting to consider partial modes with higher angular momentum as well as more realistic perturbation sources but these configurations are much more challenging to treat in massive gravity. However, even if we are not able currently to deal with such problems, we believe that they do not lead to very different results. Our opinion is supported by some calculations we have achieved by replacing the massive spin-$2$ field with the massive scalar field. Indeed, in this context and when we consider partial modes with higher angular momentum, we can observe results rather similar to those of Sec.~\ref{Sec.II}:

\begin{enumerate}[label=(\roman*)]

\item If we still describe the distortion of the Schwarzschild BH by an initial value problem \cite{DFOEH2016}.

\item If we consider the excitation of the BH by a particle plunging from slightly below the innermost stable circular orbit into the Schwarzschild BH, i.e., if we use the toy model we developed in Ref.~\cite{Decanini:2015yba} (see Figs.~\ref{fig:B_Remu_20} and \ref{fig:Reponse22mu128_FT_Adiab_Tail} and comments in figure captions).
\end{enumerate}

It would be important to extend our study to a rotating BH in massive gravity. Indeed, in that case, because the BH is described by two parameters and not just by its mass, the existence of the resonant behavior of the quasinormal excitation factors might not be accompanied by the neutralization of the associated giant ringings.

%\begin{widetext}
%\end{widetext}

We would like to conclude with some remarks inspired by our recent articles \cite{Decanini:2014bwa,Decanini:2014kha,Decanini:2015yba} as well as by the present work. The topic of classical radiation from BHs when massive fields are involved has been the subject of a large number of studies since the 1970s but, in general, they focus on very particular aspects such as the numerical determination of the quasinormal frequencies, the excitation of the corresponding resonant modes, the numerical determination of QBS complex frequencies, their role in the context of BH instability, the behavior of the late-time tail of the signal due to a BH perturbation \dots and, moreover, they consider these aspects rather independently of each other. When addressing the problem of the construction of the waveform generated by an arbitrary BH perturbation and its physical interpretation, these various aspects must be considered together and this greatly complicates the task. If we work in the low-mass regime, its seems that, {\it mutatis mutandis}, the lessons we have learned from massless fields provide a good guideline but, if this is not the case, we face numerous difficulties. It is possible to overcome the numerical difficulties encountered (see Sec.~\ref{Sec.IIB1}) but, from the theoretical point of view, the situation is much more tricky and, in particular, the unambiguous identification of the different contributions (the ``prompt'' contribution, the QNM and QBS contributions, the tail contribution \ldots) in waveforms or in the retarded Green function is not so easy and natural as for massless fields. In fact, it would be interesting to extend rigorously, for massive fields, the nice work of Leaver in Ref.~\cite{Leaver:1986gd} but, in our opinion, due to the structure of the Riemann surfaces involved as well as to the presence of the cuts associated with the wave number $p(\omega)$ [see Eq.~(\ref{p_omega})] and with the function $[\omega/p(\omega)]^{1/2}$ [see, e.g., in Eqs.~(\ref{bc_2_in}) and (\ref{bc_1_up})], this is far from obvious and certainly requires uniform asymptotic techniques.

\section{Acknowledgments}
We wish to thank Andrei Belokogne for various discussions and the ``Collectivit\'e Territoriale de Corse" for its support through the COMPA project.

\bibliography{Giant_BH_ringing}

\end{document}